\documentclass[12pt, preprint]{aastex}
\usepackage[dvips]{color}

\def\ltsima{$\;\buildrel < \over \sim \;$}
\def\simlt{\lower.5ex \hbox{\ltsima}}
\def\gtsima{$\;\buildrel > \over \sim \;$}
\def\simgt{\lower.5ex \hbox{\gtsima}}

\shorttitle{Search for H$_2^{17}$O and H$_2^{18}$O in IRC+10216}
\shortauthors{Neufeld et al.}

\begin{document}

\title{{\it Herschel}/HIFI search for H$_2^{17}$O and H$_2^{18}$O in IRC+10216: \\
constraints on models for the origin of water vapor}
\author{David A.~Neufeld\altaffilmark{1}, Volker~Tolls\altaffilmark{2},
Marcelino~Ag\'undez\altaffilmark{3,4}, 
Eduardo~Gonz\'alez-Alfonso\altaffilmark{5}, Leen~Decin\altaffilmark{6,7}, Fabien~Daniel\altaffilmark{8},
Jos\'e~Cernicharo\altaffilmark{8}, Gary~J.~Melnick\altaffilmark{2}, 
Miroslaw~Schmidt\altaffilmark{9} and Ryszard~Szczerba\altaffilmark{9}}
\altaffiltext{*}{Herschel is an ESA space observatory with science instruments provided
by European-led Principal Investigator consortia and with important
participation from NASA}
\altaffiltext{1}{Department of Physics and Astronomy, Johns Hopkins University,
3400~North~Charles~Street, Baltimore, MD 21218, USA}
\altaffiltext{2}{Harvard-Smithsonian CfA, 60 Garden Street, Cambridge, MA 02138, USA}
\altaffiltext{3}{Univ. Bordeaux, LAB, UMR 5804, F-33270, Floirac, France}
\altaffiltext{4}{CNRS, LAB, UMR 5804, F-33270, Floirac, France}
\altaffiltext{5}{Universidad de Alcal\'a de Henares, Departamento de F\'{\i}sica
y Matem\'aticas, Campus Universitario, E-28871 Alcal\'a de Henares,
Madrid, Spain}
\altaffiltext{6}{Instituut voor Sterrenkunde, Katholieke Universiteit Leuven, Celestijnenlaan 200D, 3001 Leuven, Belgium} \altaffiltext{7}{Sterrenkundig Instituut Anton Pannekoek, University of Amsterdam, Science Park 904, NL-1098 Amsterdam, The Netherlands}
\altaffiltext{8}{Departamento de Astrof\'isica, Centro de Astrobiolog\'ia, CSIC-INTA, Ctra. de Torrej\'on a Ajalvir km 4, 28850, Madrid, Spain}
\altaffiltext{9}{Nicolaus Copernicus Astronomical Center, Polish Academy of Sciences, Rabia\'nska 8, 87-100, Toru\'n, Poland}

\begin{abstract}

We report the results of a sensitive search for the minor isotopologues of water, 
H$_2^{17}$O and H$_2^{18}$O, toward the carbon-rich AGB star IRC+10216 (a.k.a.\ CW Leonis)
using the HIFI instrument on the {\it Herschel Space Observatory}.
This search was motivated by the fact that any detection of isotopic enhancement in the H$_2^{17}$O and H$_2^{18}$O abundances would have strongly implicated CO photodissociation as the source of the atomic oxygen needed to produce water in a carbon-rich circumstellar envelope.  Our observations place an upper limit of 1/470 on the H$_2^{17}$O/H$_2^{16}$O abundance ratio.  Given the isotopic $^{17}$O/$^{16}$O ratio of {1/840} { inferred previously} for the photosphere of IRC+10216, this result places an upper limit of a factor $1.8$ on the extent of any isotope-selective enhancement of H$_2^{17}$O in the circumstellar material, and provides an important constraint on any model that invokes CO photodissociation as the source of O for H$_2$O production.  In the context of the clumpy photodissociation model proposed { previously} for the origin of water in IRC+10216, our limit implies that $\rm ^{12}C^{16}O $ (not $\rm ^{13}C^{16}O$ or SiO) must be the dominant source of $\rm ^{16}O$ for H$_2$O production, and that the effects of self-shielding can only have reduced the $\rm ^{12}C^{16}O $ photodissociation rate by at most a factor $\sim 2$.

\end{abstract}


\section{Introduction}

A key {\it Herschel} result of relevance to evolved stars has been the discovery of water vapor in the warm inner envelope of the carbon-rich AGB star IRC+10216 (a.k.a. CW Leonis).  Here, SPIRE, PACS, and HIFI observations of multiple water transitions emitted by the dense outflowing envelope of this star have established (Decin et al. 2010a; Neufeld et al. 2011a) the presence of warm water vapor within a few stellar radii of the stellar photosphere.  The presence of water vapor so close to the star definitively rules out a previous suggestion that the origin of the water vapor, originally detected by means of Submillimeter Wave Astronomy Satellite (SWAS) observations of a single water transition (Melnick et al. 2001, Ford \& Neufeld 2001), was the vaporization of a Kuiper Belt analog.  In addition, and very strikingly, a small HIFI survey for water vapor in eight additional carbon-rich AGB stars has led to the detection of water emission from all eight sources, suggesting that the presence of water in carbon-rich AGB stars is nearly universal (Neufeld et al.\ 2011b).  Moreover, strong similarities in all eight sources between the spectral line profiles of water and those of other species such as CO argue against the water being released from a flattened structure such as a Kuiper Belt analog.

The widespread occurence of water in these sources is surprising, because the carbon-to-oxygen ratio is the critical determinant of the photospheric chemistry in evolved stars.  The photospheres of oxygen rich-stars, with C/O ratios $<$ 1, are dominated by CO and H$_2$O; those of carbon-rich stars, by contrast, are dominated by CO, HCN, and C$_2$H$_2$  and -- under conditions of thermochemical equilibrium -- are expected to contain very little H$_2$O.  The water abundances derived from Herschel observations of carbon-rich AGB stars are typically 3 to 4 orders of magnitude larger than
the photospheric abundance expected under conditions of thermochemical equilibrium.
This huge discrepancy, first revealed in IRC+10216 by SWAS, had led to the suggestion of several possible
origins for the water vapor, including (1) the vaporisation of icy objects (comets or dwarf planets) in orbit around the star (Ford \& Neufeld 2001); (2) Fischer-Tropsch catalysis (Willacy 2004); (3) photochemistry within an outer, photodissociated shell (Ag\'undez \& Cernicharo 2006); (4) photochemistry within a clumpy outflow (Decin et al.\ 2010a; Ag\'undez, Cernicharo \& Gu{\'e}lin 2010); (5) non-equilibrium chemistry associated with pulsationally-driven shock waves (Cherchneff 2011, 2012).  At least in the case of IRC+10216, the first three of these suggestions can been ruled out by the relative strengths of the many water transitions detected by {\it Herschel.}  Here, the large relative strength of high-lying water transitions indicates the presence of warm water vapor close to the star, whereas the models for origins (1) - (3) predict an absence of abundant water within $\sim $ 100 AU of the star. 

Using the Heterodyne Instrument for the Infrared (HIFI; de~Graauw et al.\ 2010) on board the {\it Herschel Space Observatory} (Pilbratt et al.\ 2010) , we have attempted to distinguish between the remaining origins (4 and 5 above) by means of deep searches for the minor isotopologues H$_2^{17}$O and  H$_2^{18}$O.  Any explanation for the presence of water vapor in a carbon-rich environment must posit the release of oxygen atoms from the strongly-bound CO molecule.  If that release results from photodissociation by ultraviolet radiation, an enhancement in the abundances H$_2^{17}$O and  H$_2^{18}$O might be expected relative to that of H$_2^{16}$O.  This possibility follows from the physics of CO photodissociation, which takes place following {\it line} absorption to predissociated electronic states.  Therefore,  once the transitions of relevance become optically-thick, CO can shield itself from photodissociation; thus the photodissociation rate for $\rm ^{13}CO$, $\rm C^{17}O$ and $\rm C^{18}O$ (per molecule) can significantly exceed that of $\rm ^{12}C^{16}O$.   Similar effects have been proposed, for example, by Lyons \& Young (2005) as an explanation for isotopic anomalies in the solar nebula. 

This isotope-selective photodissociation of CO preferentially produces $\rm ^{17}O$ and $\rm ^{18}O$, which will then go on to form the minor isotopologues of water vapor.    By contrast, an origin for the water vapor that does {\it not} involve the photodissociation of CO, as in the shock-driven chemistry model proposed very recently by Cherchneff (2011, 2012), makes the strong prediction that the H$_2^{17}$O/H$_2^{16}$O and H$_2^{18}$O/H$_2^{16}$O ratios should equal the isotopic abundance ratios in the photosphere: $\rm ^{17}O/^{16}O = 1/840$ and $\rm ^{18}O/^{16}O \sim 1/1260$ (Kahane et al.\ { 1992})\footnote{ These isotopic abundance ratios were determined from single-dish observations of optically-thin transitions of $\rm ^{13}CO$, $\rm C^{17}O$, $\rm C^{18}O$, $\rm ^{13}CS$, and $\rm C^{34}S$ within the circumstellar outflow, under the assumption that isotopic fractionation and isotope-selective photodissociation are negligible for these minor isotopologues of CO and CS.  Kahane et al.\ ({ 1992}) justified that assumption with reference to theoretical models -- which suggest that fractionation is only important in the outermost part of the envelope -- and by comparing the observed line profiles, which provide no evidence for any radial dependence in the relative abundances of the various isotopologues.}  This prediction follows from the fact that gas-phase reactions are unlikely to provide any significant isotopic fractionation, because the particle kinetic energies at the high temperatures of the inner envelope are much larger than any zero-point energy differences between the relevant molecular isotopologues.

As discussed by Neufeld et al.\ (2011a), previous data available from a full HIFI spectral scan of IRC+10216 place upper limits of $5 \times 10^{-3}$ ($3 \sigma$) on the H$_2^{17}$O/H$_2^{16}$O and H$_2^{18}$O/H$_2^{16}$O ratios.  The observations obtained from the full HIFI spectral survey of IRC+10216 necessarily entail relatively short integration times at any given frequency within the large HIFI frequency range.  In this paper, we report results of significantly improved sensitivity obtained from  deep observations that target the $1_{10}-1_{01}$ transitions of H$_2^{17}$O at 552.021~GHz and of H$_2^{18}$O at 547.676~GHz.   The H$_2^{16}$O $1_{10}-1_{01}$ line at 556.936~GHz had been observed previously by {\it Herschel}/HIFI but -- to ensure a meaningful comparison of lines measured toward a variable star -- was reobserved at the same time as the H$_2^{17}$O and H$_2^{18}$O transitions.
In \S 2 below, we discuss the new observations and the methods used to reduce the data.  The results are presented in \S 3 and discussed in \S 4.

\section{Observations and data reduction}

The observations of IRC+10216 were carried out on 2011 May 18 in the Open Time program OT2$\_$dneufeld$\_$6.  We used HIFI in dual beam switch (DBS) mode to target the $1_{10}-1_{01}$ rotational transitions of H$_2^{16}$O, H$_2^{17}$O, and H$_2^{18}$O in mixer band 1a.  The details of each observation are given in Table 1.    The telescope beam, of diameter $\sim 38^{\prime\prime}$ (HPBW), was centered on IRC+10216 at coordinates $\alpha=\rm 9h\, 47m \, 57.41s$, $\delta= +13^0 16^\prime 43.6^{\prime \prime}$ (J2000), and the reference positions were located at offsets of 3$^\prime$ on either side of the source.  The wide band spectrometer (WBS) was used to obtain a spectral resolution of 1.1~MHz, corresponding to a Doppler velocity $\sim 0.6$~km/s at the frequency of the observed transitions.  The data were processed using the {\it Herschel} Interactive Processing Environment (HIPE; Ott 2010), version 9.0.0, providing fully calibrated spectra with the intensities expressed as antenna temperature and the frequencies in the frame of the Local Standard of Rest (LSR).  Given a {\it Herschel} aperture efficiency of 0.68 (Roelfsema et al.\ 2012), the ratio of antenna temperature to flux is 2.1 mK/Jy for an unresolved source at the center of the beam.

For each spectral line, we used three separate local oscillator (LO) frequencies, evenly spaced by a small offset, to facilitate sideband deconvolution.  The standard deconvolution tool is optimized for spectral scans in which the redundancy is 4 or greater and the LO spacings are slightly uneven; if the redundancy is less than 4, the standard method fails to remove small artifacts. 
We therefore used a maximum-entropy deconvolution method to decompose the double sideband spectra into a pair of single-sideband spectra. Because the entropy of the artifacts mentioned above is small, this method allows a reliable solution to be determined by minimizing $\chi^2$ while keeping the so-called ``gain entropy'' high. 
We were able to confirm the results of this method by comparing them with spectra derived by manually separating the strongest LSB and USB lines. 

After performing the sideband deconvolution as described above, we found the resultant spectra obtained for the horizontal and vertical polarizations to be very similar in their appearance and noise characteristics.  We therefore coaveraged the two orthogonal polarizations, and -- to improve the signal-to-noise ratio -- then rebinned the spectra to a velocity-resolution of 3 km/s.

\section{Results}

Figures 1 and 2 show the spectra targeting H$_2^{17}$O and H$_2^{18}$O obtained from the data reduction procedure described in \S2 above. 
The underlying noise $\sim 1$~mK (r.m.s. in the smoothed spectra) is consistent with the predictions of the HSPOT time estimator, but the spectra are clearly characterized by a high density of emission features.  For lines with peak antenna temperatures exceeding $\sim 10$~mK ($\sim 5$~Jy), plausible identifications were readily obtained from standard catalogs\footnote{The Cologne Database for Molecular Spectroscopy (CDMS; M\"uller et al. 2001, 2005), the JPL line list (Pickett et al.\ 1998), and Splatalogue (Remijan \& Markwick-Kemper 2007)} and are marked in Figures 1 and 2; here the width of each mark indicates the uncertainty in the line frequency (or, in the case of AlCl, the frequency spread of the hyperfine components).  However, for lines weaker than $\sim 5$~Jy, no plausible identification was obvious.  Many of these weak unidentified lines are relatively narrow, implying expansion velocities $\le 6$~km/s and suggesting an origin close to the star.  In this respect, the spectrum is rather similar to that obtained in the 345 GHz spectral survey of IRC+10216 carried out by Patel et al.\ (2011), which revealed a new population of weak, narrow, and mainly unidentified emission features that the authors attributed to uncatalogued transitions within vibrationally-excited states.  Given the density of emission features in the spectra that we obtained, the incompleteness of current line catalogs unfortunately limits our ability (more so than the intrinsic signal-to-noise ratio) to identify very weak emissions from H$_2^{17}$O and H$_2^{18}$O unequivocally.  (On the other hand, improvements in the limits derived below may be possible as line catalogs become more complete, allowing weak interlopers to be identified and their relative line strengths modeled.)

In Figures 3 and 4, we show the spectra targeting H$_2^{17}$O and H$_2^{18}$O (black histograms) on an expanded velocity scale, with vertical dotted lines indicating the velocity centroid (--26.5~km/s) and outflow velocity (14.5~km/s) inferred from observations of other spectral lines.  For the H$_2^{17}$O transition (Figure 3), there is no evidence for excess emission at the LSR velocity of the source, although our sensitivity is clearly limited by unidentified emission features at nearby velocities.  The magenta histogram in this figure shows the H$_2^{16}$O line scaled by a factor of 1/220; this being the strongest feature that could be accommodated under the observed spectral profile, we adopt 1/220 as an upper limit on the H$_2^{17}$O/H$_2^{16}$O line ratio.  { We note here that the H$_2^{16}$O line profile is considerably broader than the narrow unidentified emission features discussed above; its breadth is typical of other transitions within the ground vibrational states of circumstellar molecules, and is consistent with the predictions of excitation models for the H$_2^{16}$O $1_{10}-1_{01}$ line (e.g. Gonz{\'a}lez-Alfonso, Neufeld, \& Melnick 2007).}  

For the H$_2^{18}$O spectrum (Figure 4), a feature is clearly apparent close to the expected frequency of the H$_2^{18}$O transition, with an integrated antenna temperature $\sim 71$~mK~km/s that exceeds our upper limit on the H$_2^{17}$O line.  Given the $^{17}$O/$^{18}$O isotopic ratio $\sim 1.5$ in IRC+10216 (Kahane et al.\ { 1992}), it would be surprising to find a $1_{10}-1_{01}$ line of H$_2^{18}$O that was {\it stronger} than that of H$_2^{17}$O.  A careful search of available line catalogs revealed an interloper transition of vibrationally-excited HNC, the $\nu_2=1f$ $J=6-5$ line, which has a frequency offset (relative to H$_2^{17}$O $1_{10}-1_{01}$) of only $\sim - 6 \, \rm MHz$, corresponding to a velocity offset $\sim +3$~km/s.  Based upon a model for the excitation of HNC, constrained by {\it Herschel} observations of the ground vibrational state (Daniel et al.\ 2012) along with ALMA observations of transitions of lower frequency within the $\nu_2=1$ state (Daniel et al.\, in preparation), we expect a velocity integrated antenna temperature of 62~mK~km/s for this HNC transition (with the line strength and profile shown in red).  After subtraction of the expected HNC emission, the observed spectrum (blue) shows no clear evidence for any residual that could be attributed to H$_2^{18}$O.  

\section{Discussion}

Because our sensitivity to H$_2^{18}$O is limited by the precision with which we can model the interloper HNC $\nu_2=1f$ $J=6-5$ line, and given the { inferred} {\it photospheric} isotopic ratio of $^{17}$O/$^{18}$O $\sim 1.5$, we will focus on the H$_2^{17}$O line intensity limit as a constraint on the extent of any isotope-selective enhancement in the minor isotologues of water.  
Using 
%two independent excitation models for the excitation of water in IRC+10216, the GASTRoNOoM model (Decin et al.\ 2006)
the model of Gonz{\'a}lez-Alfonso et al.\ (2007) for the excitation of water in IRC+10216, we found that the observed line intensity limit, H$_2^{17}$O $1_{10}-1_{01}$/ H$_2^{16}$O $1_{10}-1_{01} < 1/220$ implies an abundance ratio limit H$_2^{17}$O / H$_2^{16}$O $< 1/470$ (the latter being smaller than the former on account of optical-depth effects for H$_2^{16}$O).  Comparing this with the isotopic $^{17}$O/$^{16}$O ratio of 1/840, we can place a conservative upper limit of 1.8 on the factor, $f_{\rm e}$, by which the minor isotopologues are enhanced.  

This limit on $f_{\rm e}$ places a strong constraint on any model in which the photodissociation of CO is the source of atomic oxygen to form water in IRC+10216.  For example, were $^{13}$CO the primary source of $^{16}$O {and $\rm C^{17}O$ that of $^{17}$O}, then the expected enhancement factor (Neufeld et al.\ 2011a) would be $f_{\rm e} \sim$~$\rm ^{12}C/^{13}C =45$ (Cernicharo et al.\ 2000).  Such a large enhancement factor is entirely inconsistent with the data reported here, and would argue against the clumpy photodissociation model and in favor of the alternative model proposed by Cherchneff (2011, 2012).  
However, as noted by Neufeld et al.\ { 2011a}), this large predicted enhancement might be decreased if photodissociation of $\rm ^{12}CO$ or SiO contributed significantly to $^{16}$O production.  Moreover, the photodissociation of $\rm ^{12}CO$ in vibrationally-excited states that are populated close to the star might further reduce the enhancement factor. A proper account of these effects will require an extensive theoretical study, beyond the scope of this {\it Letter}, in which the clumpy photodissociation model is investigated fully with detailed treatments of radiative transfer and of the photodissociation of vibrationally-excited CO.

{In any such model, photodissociation of $\rm ^{12}C^{16}O$ would necessarily have to be the primary source of $\rm ^{16}O$.  Barring any gas-phase fractionation effects, which would to have be extremely minor at the high gas temperatures in the inner envelope, the observed upper limit on $\rm H_2^{16}O$/$\rm H_2^{17}O$ would imply that $\rm ^{16}O$ must be produced at least 470 times as fast (per unit volume) as $\rm ^{17}O$.  Since the latter is produced by photodissociation of $\rm ^{12}C^{17}O$, the  photodissociation of $\rm ^{12}C^{16}O$ is the only photodissociation process that could possibly provide $\rm ^{16}O$ at a sufficient rate; no other oxygen-bearing molecule is abundant enough to yield the required  $\rm ^{16}O$ production rate as a result of its photodissociation.\footnote{Although SiO has an unshielded photodissociation rate that is $\sim 6$ times as large as that of $\rm ^{12}C^{17}O$ (van~Dishoeck, Jonkeid \& van~Hemert\ 2006; Visser, van~Dishoeck \& Black 2009), its abundance is only $\sim \rm few \times 10^{-7}$ relative to H$_2$ (Decin et al.\ 2010b), i.e.\ even less than that of $\rm ^{12}C^{17}O$.}}

\acknowledgments

HIFI has been designed and built by a consortium of institutes and university departments from across
Europe, Canada and the United States under the leadership of SRON Netherlands Institute for Space
Research, Groningen, The Netherlands and with major contributions from Germany, France and the US.
Consortium members are: Canada: CSA, U.~Waterloo; France: CESR, LAB, LERMA, IRAM; Germany:
KOSMA, MPIfR, MPS; Ireland, NUI Maynooth; Italy: ASI, IFSI-INAF, Osservatorio Astrofisico di Arcetri-
INAF; Netherlands: SRON, TUD; Poland: CAMK, CBK; Spain: Observatorio Astron\'omico Nacional (IGN),
Centro de Astrobiolog\'ia (CSIC-INTA). Sweden: Chalmers University of Technology - MC2, RSS \& GARD;
Onsala Space Observatory; Swedish National Space Board, Stockholm University - Stockholm Observatory;
Switzerland: ETH Zurich, FHNW; USA: Caltech, JPL, NHSC.

This research was performed, in part, through a JPL contract funded by the National Aeronautics and Space Administration.
E.G-A is a Research Associate at the Harvard-Smithsonian
Center for Astrophysics, and thanks the Spanish 
Ministerio de Ciencia e Innovaci\'on for support under project AYA2010-21697-C05-01.
%E.G-A  is a Research Associate at the Harvard-Smithsonian 
%Center for Astrophysics.
%R.~Sz. and M.~Sch.\ acknowledge support from grant N 203 393334.   
%This work has been partially supported by the Spanish MICINN, program CONSOLIDER INGENIO 2010, grant ASTROMOL (CSD2009-00038).
R.\ Sz.\ and M.\ Sch.\ acknowledge support from Polish NCN grant no. 203 581040.

\begin{deluxetable}{cccc}

\tablewidth{0pt}
\tablecaption{Observations of IRC+10216} 
\tablehead{Target  & $\nu$  & $t_{\rm obs}^a$ & AOR $\#$s  \\
           transition            & (GHz) & (s) &  }

\startdata
H$_2^{18}$O $1_{10}-1_{01}$ & 547.676 & 29701 & 1342246477--79 \\
H$_2^{17}$O $1_{10}-1_{01}$ & 552.021 & 13104 & 1342246480--82 \\
H$_2^{16}$O $1_{10}-1_{01}$ & 556.936 & \phantom{00}434 & 1342246483--85 \\

\enddata
\tablenotetext{a}{Total observing time for three LO settings}
%\tablenotetext{b}{R.m.s noise in the sideband-deconvolved, rebinned, polarization-averaged spectra (see text)}
\end{deluxetable}

\begin{figure}
\includegraphics[width=15 cm]{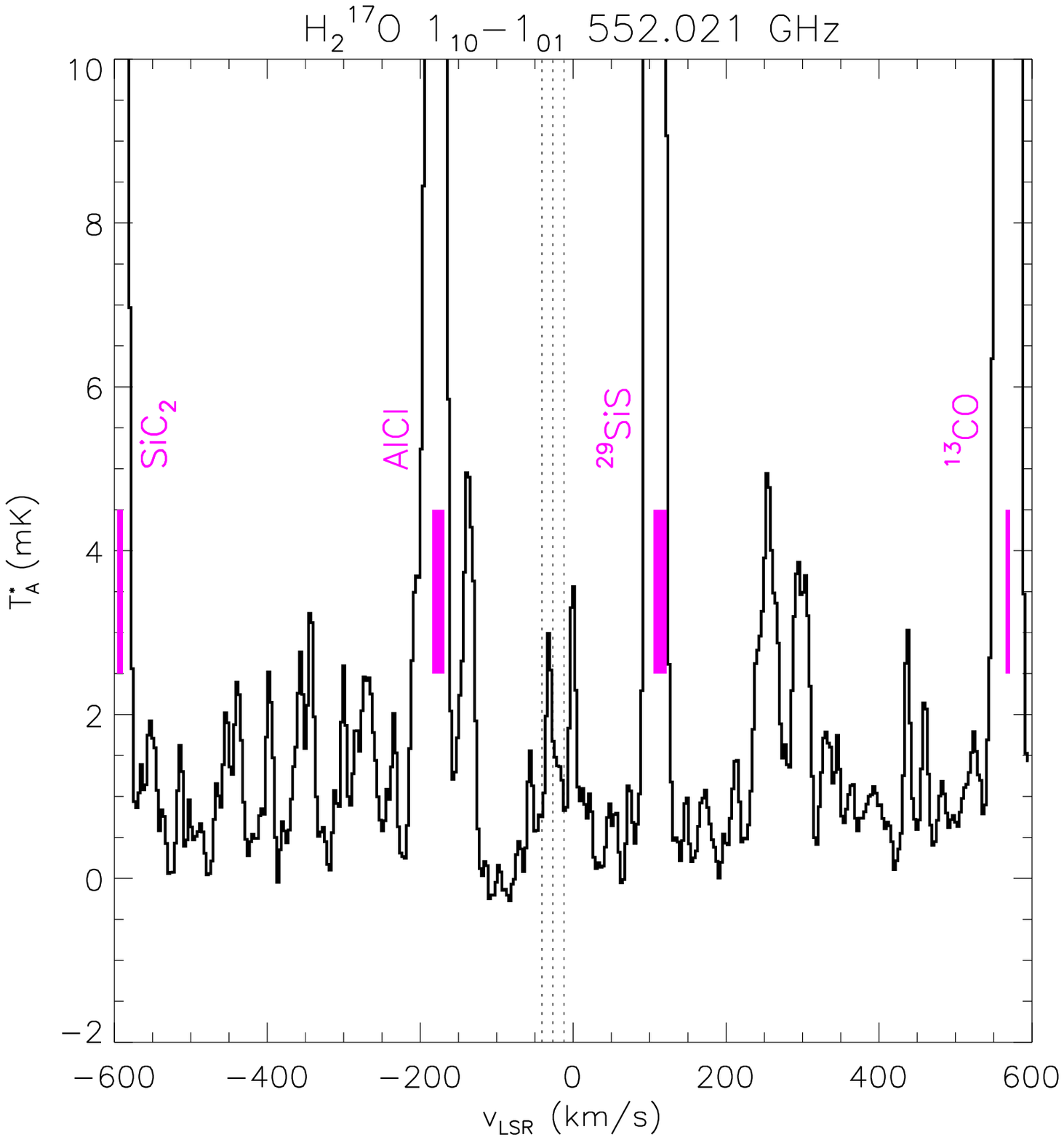}
\caption{Spectrum targeting the $1_{10}-1_{01}$ transition of H$_2^{17}$O in IRC+10216.  The LSR velocity scale is for the 552.021~GHz rest frequency of that transition.  Vertical dotted lines indicate the velocity centroid (--26.5 km/s) and outflow velocity (14.5 km/s) determined from observations of other lines.  Purple marks indicate probable line identifications obtained from standard catalogues (see text), with the width of each mark indicating the uncertainty in the line frequency, { or - in the case of AlCl - the frequency spread of the hyperfine components}.}
\end{figure}

\begin{figure}
\includegraphics[width=15 cm]{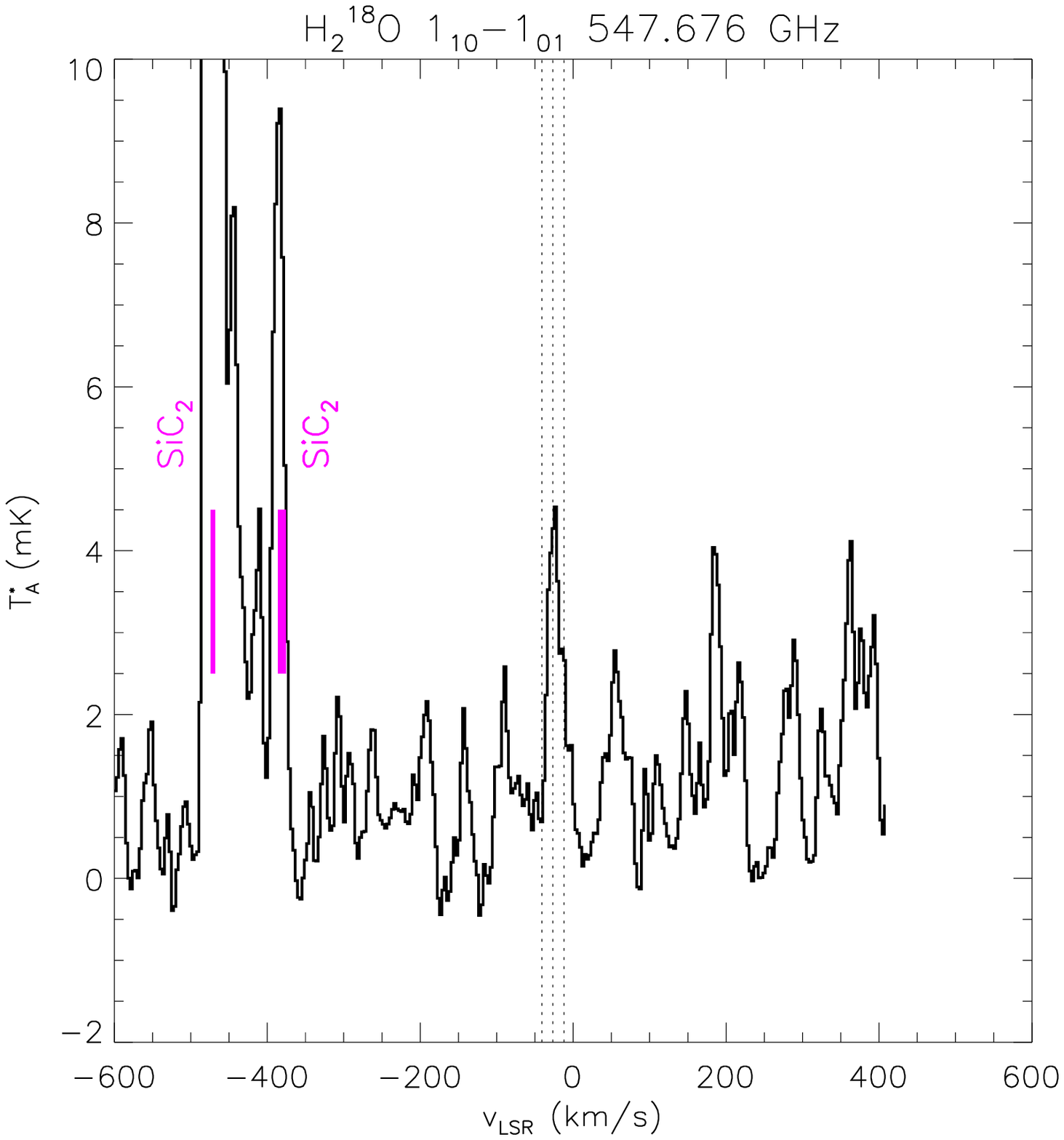}
\caption{Spectrum targeting the $1_{10}-1_{01}$ transition of H$_2^{18}$O in IRC+10216.  The LSR velocity scale is for the 547.676~GHz rest frequency of that transition.  Vertical dotted lines indicate the velocity centroid (--26.5 km/s) and outflow velocity (14.5 km/s) determined from observations of other lines.  Purple marks indicate probable line identifications obtained from standard catalogues (see text), with the width of each mark indicating the uncertainty in the line frequency, { or - in the case of AlCl - the frequency spread of the hyperfine components}.}
\end{figure}

\begin{figure}
\includegraphics[width=15 cm]{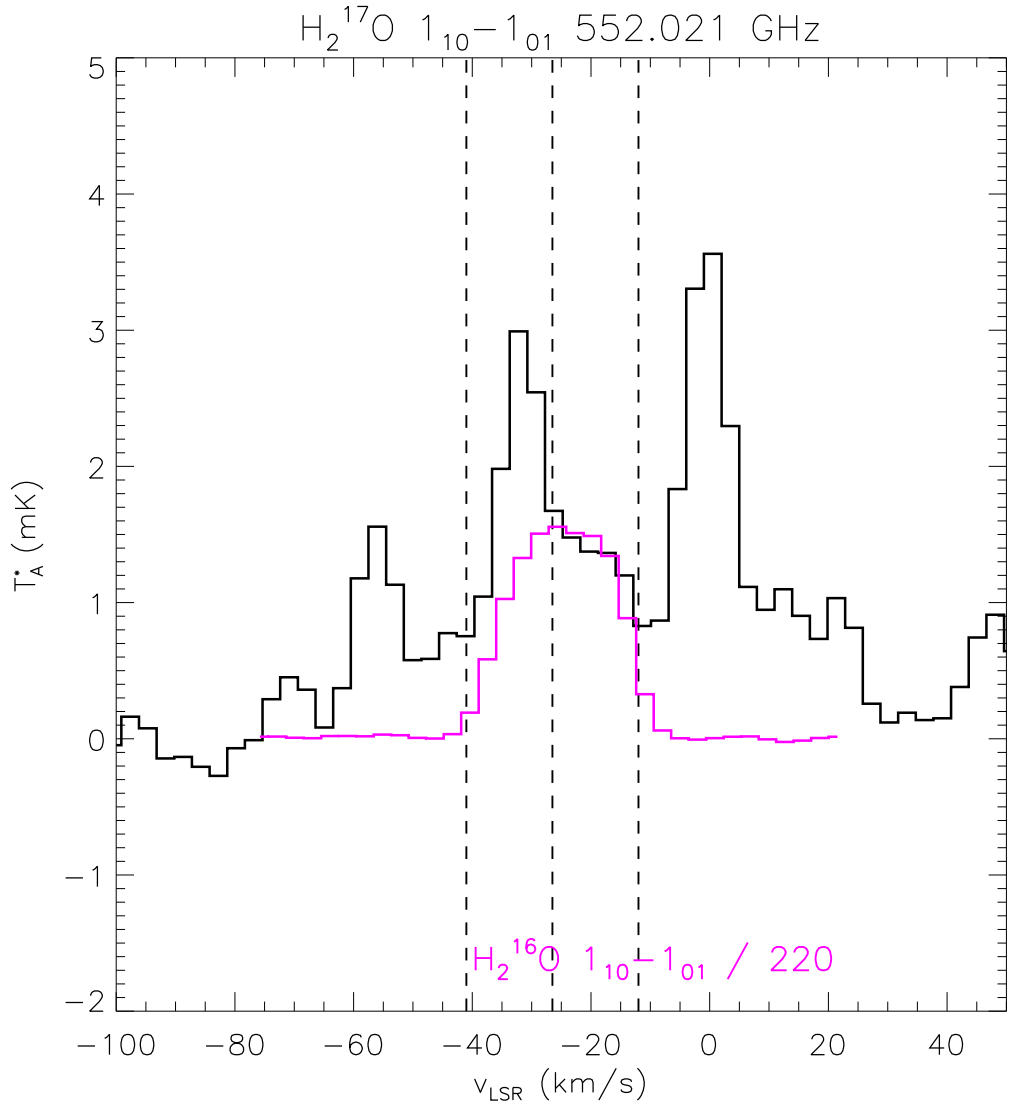}
\caption{Same as Figure 1, but on an expanded velocity scale. Magenta histogram: H$_2^{16}$O  $1_{10}-1_{01}$ line scaled by a factor of 1/220.}
\end{figure}

\begin{figure}
\includegraphics[width=15 cm]{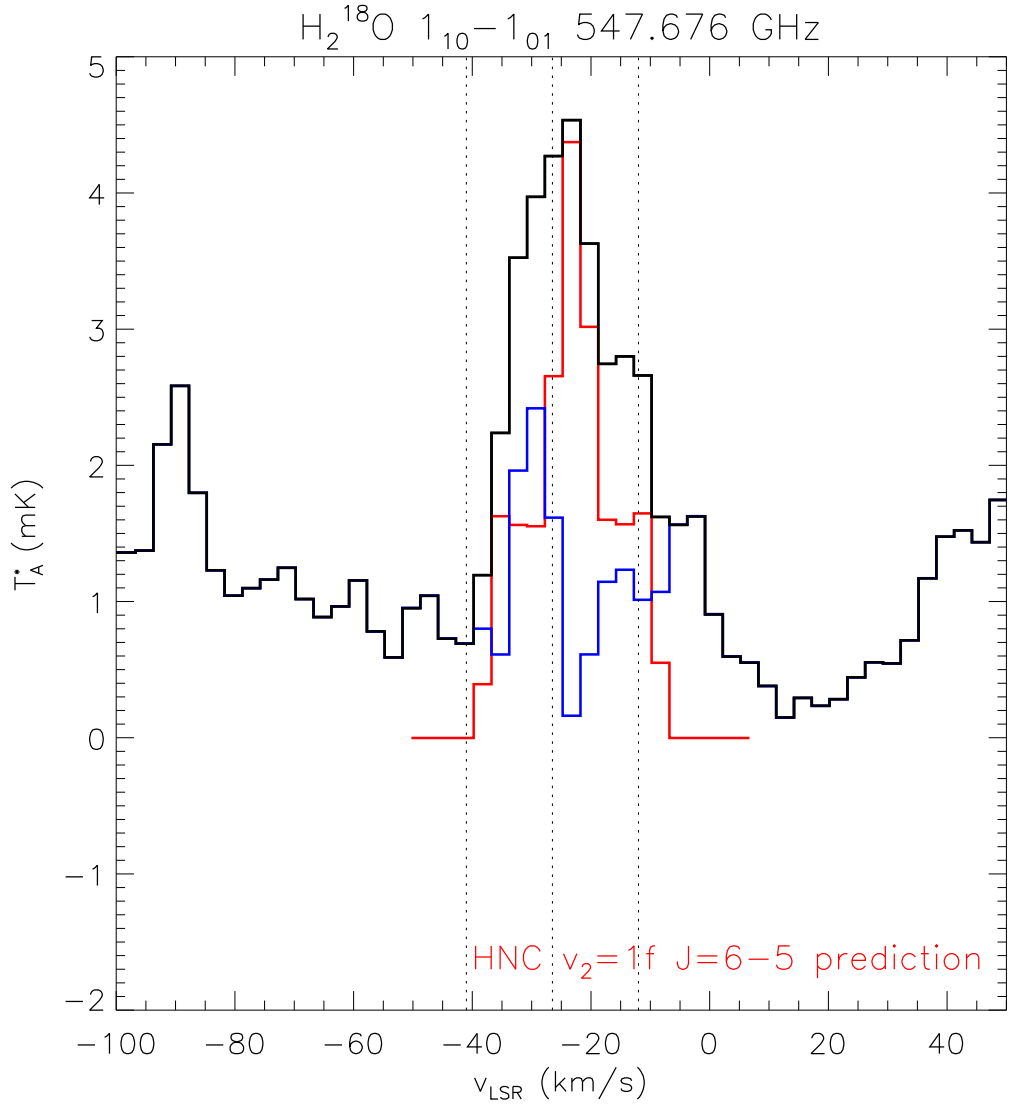}
\caption{Same as Figure 2, but on an expanded velocity scale. Red histogram : prediction for the HNC $\nu_2=1f$ $J=6-5$ line.  Blue histogram: data, after subtraction of the emission attributable to HNC $\nu_2=1f$ $J=6-5$.}
\end{figure}
\end{document}